\RequirePackage{fix-cm}
\documentclass[epjc3]{svjour3}  
\smartqed 
\RequirePackage{graphicx}
\RequirePackage{mathptmx}      
\RequirePackage{latexsym}
\RequirePackage[numbers,sort&compress]{natbib}
\RequirePackage[colorlinks,citecolor=blue,urlcolor=blue,linkcolor=blue]{hyperref}
\usepackage[utf8]{inputenc}
\usepackage{amsmath,amssymb,amsbsy,amstext,amsfonts}
\usepackage{graphicx,multicol}
\usepackage[svgnames,dvipsnames,x11names]{xcolor}
\usepackage{hyperref}
\usepackage{extarrows}

\journalname{Eur. Phys. J. C}

\begin{document}

\title{\textbf{Islands in Kaluza-Klein black holes}
}

\author{Yizhou Lu\thanksref{e1,addr1}
        \and
        Jiong Lin\thanksref{e2,addr1} 
}

\thankstext{e1}{e-mail: louischou@hust.edu.cn}
\thankstext{e2}{e-mail: jionglin@hust.edu.cn (Corresponding author)}


\institute{School of Physics, Huazhong University of Science and Technology,\\Wuhan, Hubei 430074, China \label{addr1}
}

\date{Received: date / Accepted: date}

\maketitle

\begin{abstract}
The newly proposed island formula for entanglement entropy of Hawking radiation is applied to spherically symmetric 4-dimensional eternal Kaluza-Klein (KK) black hole. 
The "charge" $Q$ of KK black holes quantifies its deviation from Schwarzschild black holes.
The impact of $Q$ on the island is studied at late times.
The late-time island, whose boundary is located outside but within a Planckian distance of the horizon, is slightly extended by $Q$.
While the no-island entropy grows linearly, the late-time entanglement entropy is given by island configuration with twice the Bekenstein-Hawking entropy.
Thus we reproduce the Page curve for the eternal KK black holes. 
Compared with Schwarzschild results, the Page time is delayed by a factor $(1+Q/r_h)$ and the scrambling time is prolonged by a factor $(1+Q/r_h)^{1/2}$.
Moreover, the higher-dimensional generalization is presented. 
Skeptically, there are Planck length scales involved, in which a semi-classical description may break down.
\end{abstract}

\section{Introduction}
\label{intro}
The pursuit of a quantum theory of gravity is one of the most important tasks in modern physics while it remains mysterious.
A nice object, both of theoretical and observational interests, for studying the quantum effect of gravity is the black hole. 
Black holes have thermodynamics. 
When coupled to a quantum field theory, a black hole can emit Hawking radiation
\cite{Hawking:1974sw,Hawking:1976ra} at Hawking temperature, and its entropy is proportional to the area of the horizon. 
If black holes form via gravitational collapse from a pure state and then start to radiate, the entanglement entropy of Hawking radiation will grow from zero at the beginning.
According to Hawking's calculation, the entanglement entropy of the radiation is constantly increasing.
If Hawking is right, then after the black hole evaporates, the system will be in a mixed state, which is contradictory to unitary evolution in quantum theory. 
To protect unitarity, in fact, the entanglement entropy should follow the Page curve \cite{Page:1993wv,Page:2013dx,Page:1993df}. 
Reproducing the Page curve for the entanglement entropy of Hawking radiation by explicit calculation is the key point to the information paradox. 

Recently, a significant progress has been made by introducing regions called \emph{islands} \cite{Penington:2019npb,Penington:2019kki,Almheiri:2019psf,Almheiri:2019qdq,Almheiri:2019hni} to black holes. 
Islands, as the name implies, are some disconnected regions $ I$ that actually belong to the entanglement wedge of Hawking radiation $\mathcal{R}$. 
To wit, if we are going to compute the entanglement entropy for Hawking radiation, we have to take into account the \emph{islands}. 
In this way, the Page curve is indeed recovered in simple 2D models \cite{Almheiri:2019hni,Almheiri:2019yqk}.
In the spirit of quantum extremal surface (QES) prescription \cite{Ryu:2006bv,Hubeny:2007xt,Engelhardt:2014gca}, the island formula for the entanglement entropy of the Hawking radiation is given by
\begin{equation}\label{island_formula}
S(\mathcal{R})=\mathrm{Min}_{\partial  I}\left\{
\mathrm{Ext}_{\partial I}\left[\frac{\mathrm{Area}(\partial I)}{4G_N}+S_{\mathrm{mat}}({\mathcal{R}\cup I})\right]
\right\},
\end{equation}
where we extremize the generalized entropy $S_{\mathrm{gen}}$ \cite{Lewkowycz:2013nqa} over the possible boundaries of islands and take the minimum.\footnote{We maximize $S_{\mathrm{gen}}$ over the time direction of $\partial I$.} Eq. \eqref{island_formula} can be viewed as a generalization of RT/HRT formula. For a nice conceptual review, see \cite{Almheiri:2020cfm}.

The island formula \eqref{island_formula} can also be derived from gravitational path integral \cite{Almheiri:2019qdq,Penington:2019kki}.
By using the replica trick \cite{Callan:1994py,Holzhey:1994we,Calabrese:2009qy,Casini:2009sr}, the entanglement entropy is given by 
\begin{equation}
    S=-\partial_n\left.\left(
    \frac{\log[\mathrm{Tr}(\rho^n)]}{n}
    \right)\right|_{n=1},
\end{equation}
where $\rho$ is the reduced density matrix of the region where we are computing the entropy. 
Evaluating $\mathrm{Tr}\rho^n$ involves a manifold $\widetilde{\mathcal{M}}_n$ that consists of $n$ copies of the original system sewed cyclically, where the path integral will be performed.
If there is a region of gravity, we should take into account all possible solutions to gravity (different topologies), subject to fixed boundary condition, in gravitational path integral. 
Particularly, besides simple disks (Hawking saddles), one should also consider a wormhole that connects the $n$ replica sheets on the gravity sector. This kind of saddles is called \emph{replica wormholes}. The presence of replica wormholes in replica geometry indicates the existence of entanglement islands in the original system. 
For island geometry to solve the equation of motion on the boundary of gravity, one obtains a condition that is exactly given by QES \eqref{island_formula}. 
In this sense, the QES prescription is interpreted by gravitational path integral.
According to the QES prescription, 
\begin{equation}\label{QES_S}
    S=\mathrm{min}\left\{
    S_{\mathrm{gen}}^{\mathrm{no~ island}},S_{\mathrm{gen}}^{\mathrm{island}},\cdots
    \right\},
\end{equation}
where the ellipsis denotes other configurations  that are subdominant. 
At early stage, the entanglement entropy will be given by the one sans island. 
But at late times, the island geometry always dominates so that the Page curve is recovered.
The authors of \cite{Almheiri:2019qdq} explicitly did the calculation to show this procedure in a 2-dimensional model. 
They considered a Jackiw-Teitelboim (JT) gravity living in a nearly AdS$_2$ \cite{Almheiri:2014cka} and coupled to a 2-dimensional conformal field theory (CFT$_2$). There also is a flat space filled with CFT$_2$ outside the AdS$_2$. 
This is a toy model describing near-horizon behavior of higher dimensional near-extremal Reissner-Nordstr\"{o}m (RN) black hole.
Refs. \cite{Goto:2020wnk,Hollowood:2020cou,Gautason:2020tmk} extended it by considering an evaporating black hole.

Though the explicit computation was performed in 2-dimension, island formula \eqref{island_formula} is expected to work for higher-dimensional black holes. 
Recent phenomenological works confirmed that it extends to some higher dimensional black holes \cite{Almheiri:2019psy,Hashimoto:2020cas,Wang:2021woy,Karananas:2020fwx,Kim:2021gzd,Ling:2020laa,Alishahiha:2020qza,Geng:2020qvw,Chu:2021gdb,Bak:2020enw}, as well as to some other 2-dimensional black holes \cite{Wang:2021mqq,Alishahiha:2020qza,Almheiri:2019yqk}.
In particular, a black hole in a 2-dimensional model similar to that in \cite{Almheiri:2019qdq} was studied in \cite{Almheiri:2019yqk}, which shows that the island at late times is outside the horizon. 
Later works \cite{Hashimoto:2020cas,Kim:2021gzd,Karananas:2020fwx,Wang:2021woy} reported the same findings for some other black holes.
Surprisingly, it was reported that island rule may not save the day for the information paradox of Liouville black holes \cite{Li:2021lfo}. Islands were also discussed in the cosmological scenario \cite{Krishnan:2020fer,Hartman:2020khs,Balasubramanian:2020xqf,Chen:2020tes,VanRaamsdonk:2020tlr,Geng:2021wcq}, especially in de Sitter spacetime where the cosmological constant is positive. 
For other relevant interesting works, see the non-exhaustive list  \cite{Chen:2020uac,Chen:2020hmv,Qi:2021sxb,Balasubramanian:2020coy,Balasubramanian:2021wgd,Matsuo:2020ypv,Geng:2020fxl,Geng:2021iyq}.

String theory is the most promising theory to quantize gravity. 
After dimension reduction, the gauged supergravity could be reducted to 4 dimensional Kaluza-Klein theory \cite{Kaluza:1921tu,klein1926zf}.
Thus, it is interesting and meaningful to study Kaluza-Klein theory.
In this paper, we will study the entanglement entropy of Hawking radiation in spherically symmetric Kaluza-Klein (KK) black holes \cite{Gibbons:1985ac} applying the aforementioned island formula.
The $D$-dimensional spherically symmetric KK black hole is the solution of the Kaluza-Klein theory with Lagrangian
\begin{equation}\label{KKaction}
    \mathcal{L}=\sqrt{-g}\left(
    R-\frac{1}{2}(\partial \phi)^2-\frac{1}{4} {e}^{\sqrt{2(D-1)/(D-2)}\phi}F^2
    \right),
\end{equation}
which can be also regarded as the dimensionally-reducted Lagrangian of $D+1$ dimensional Einstein-Hilbert Lagrangian after compactifying one of the spatial coordinates on a circle $\mathbb{S}^1$.
After dimension reduction, the scalar field and the vector field emerge.
Note that to obtain Einstein gravity and 
canonical kinetic term of scalar field in $D$ dimension, 
the parameters of $D+1$ dimension metric
\begin{equation}
    d\hat{s}_{D+1}^2={e}^{2\alpha\phi}d s_{D}^2+{e}^{2\beta\phi}(d z+A_{\mu}d x^{\mu})^2
\end{equation}
should be
\begin{equation}
    \alpha^2=\frac{1}{2(D-1)(D-2)},\
\beta=-(D-2)\alpha.
\end{equation}
Unlike the (non-extremal) RN black holes that possess two event horizons, spherically symmetric KK black holes have only one event horizon due to the emergence of the non-trivial scalar field \cite{Cai:2020wrp}. 
In fact, KK black holes can recover Schwarzschild spacetime in the limit of vanishing charge.
Thus it is interesting to investigate the impact of the charge on the island compared with the Schwarzschild black holes \cite{Hashimoto:2020cas}. 

In this paper, we consider an eternal KK black hole in equilibrium to a bath of CFT. 
The eternal black hole is in a Hartle-Hawking state  that is a thermofield double state (TFD), where the introduced thermofield double (left wedge) purifies the state of the right wedge \cite{Israel:1976ur,Maldacena:2001kr,Hartle:1976tp}.
The emission of Hawking radiation balances the absorption of the black hole from the bath, so the total energy of the black hole is invariant under time evolution. 
We also assume that $Q$ remains constant.
On the contrary, the entanglement entropy between outside Hawking modes and the black hole grows with time without a bound owing to the continuing exchange of particles. 
We shall see this problem is resolved by island proposal and accordingly the Page curve is reproduced.
For a distant observer, only s-wave part contributes to our calculation of von Neumann entropy of Hawking radiation.
This allows us to use a 2D CFT to effectively describe the system. 
Besides, the greybody factor and the Schwinger effect are not considered.
The information paradox for certain eternal black holes was discussed  in \cite{Almheiri:2019yqk,Hashimoto:2020cas,Wang:2021woy,Almheiri:2019qdq,Karananas:2020fwx,Maldacena:2001kr,Kim:2021gzd}. 

This paper is arranged as follows. In Sec. \ref{sec:KKBH}, we introduce the basics of KK black holes. In Sec. \ref{sec:entropy}, we evaluate the entanglement entropy of Hawking radiation from KK black holes in cases of no island and of one island.
In Sec. \ref{sec:DdimKKBH}, the generalization to higher dimensional KK black holes is presented.
In Sec. \ref{sec:page_scrambling}, we discuss the Page time and scrambling time of KK black holes. And finally we discuss and conclude this paper.

\section{Kaluza-Klein black holes}
\label{sec:KKBH}
In this section, we give a brief review of the basics of KK black holes and set up the coordinates convenient for our calculations.

The metric of a non-rotating Kaluza-Klein black hole in 4-dimensional asymptotically flat spacetime, which is a black hole solution to \eqref{KKaction}, takes the form
\begin{equation}
    d s^2=-W(r)d t^2+\frac{d r^2}{W(r)}+H^{1/2}r^2d\Omega^2,
\end{equation}
where
\begin{equation}
    W(r)=f(r)/\sqrt{H(r)},\quad f(r)=1-\frac{r_h}{r},\quad H(r)=1+\frac{Q}{r}.
\end{equation}
Here $r_h$ is the radial coordinate of horizon, which is not necessarily the mass of the black hole. 
$Q\geq 0$ is related to the charge of the black hole \cite{Liu:2012jra} and sometimes we will refer to it as "charge", but keep in mind that it is not seriously correct.
The factor $H^{1/2}(r)$ modifies the area of the horizon.
More general charged rotating KK black hole was obtained in \cite{Wu:2011zzh}.
Define a tortoise coordinate $d r_*=d r/W(r)$, such that
\begin{equation}
    d s^2=-W(r)(d t^2-d r_*^2)+H^{1/2}r^2d\Omega^2,
\end{equation}
which is in a conformally flat form in $t$-$r_*$ plane.
Sometimes, we will assume the charge is very small $\delta\equiv Q/r_h\ll 1$, and keep only the linear order in $Q$ to see the asymptotic behavior near $Q=0$.
To the linear order in $Q$, we have
\begin{align}\label{4dWfunc}
    W(r)\simeq f(r)\left(1-\frac{Q}{2r}\right).
\end{align}
The surface gravity is then
\begin{equation}
    \kappa=\left.\frac{f'}{2\sqrt{H}}\right|_{r=r_h}=\frac{1}{2r_h}\frac{1}{\sqrt{H(r_h)}},
\end{equation}
which gives the Hawking temperature of the KK black hole
\begin{equation}
    T=\frac{\kappa}{2\pi}=\frac{1}{4\pi r_h}\frac{1}{\sqrt{H(r_h)}}.
\end{equation}
And this is also the temperature of faraway thermal bath due to thermal equilibrium.

\begin{figure}
    \centering
    \includegraphics[scale=0.4]{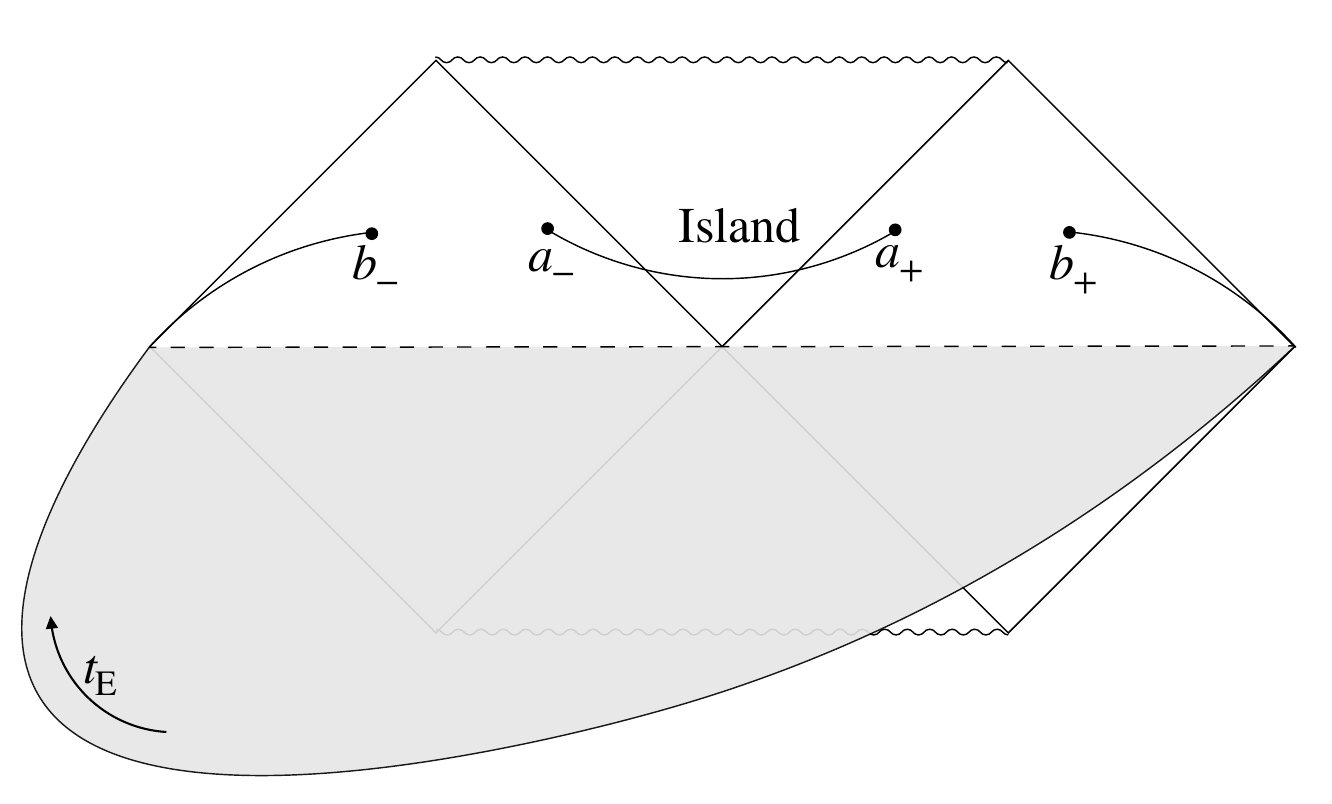}
    \caption{The Penrose diagram of an eternal Kaluza-Klein black hole resembles the Schwarzschild's. Every point on the diagram is understood to be a two-sphere. The solid curves are region of which we are computing entropy. $a_-$ and $a_+$ are the boundaries of the entanglement island. The shaded region implies that the initial state is defined by Euclidean path integral.}
    \label{fig:penrose}
\end{figure}

After explicit integration, the tortoise coordinate for KK black holes up to an integral constant is given by
\begin{equation}
    r_*(r)=\sqrt{r(Q+r)}+(Q+2r_h)\mathrm{arcsinh}\left(
    \sqrt{\frac{r}{Q}}
    \right)
    +\sqrt{r_h(Q+r_h)}\log\left|
    \frac{1-\sqrt{\frac{r}{r_h}\frac{Q+r_h}{Q+r}}}{1+\sqrt{\frac{r}{r_h}\frac{Q+r_h}{Q+r}}}
    \right|,
\end{equation}
which to the linear order in $Q$ is
\begin{equation}\label{rs1stQ}
    r_*(r)\approx r+(r_h+Q/2)\log\left|
    \frac{r-r_h}{r_h}\right|,
\end{equation}
up to an integral constant.  
We can define new coordinates to write the metric as
\begin{equation}\label{kruskal}
    d s^2=-W(r)e^{-2\kappa r_*(r)}d X^+d X^-+H^{1/2}r^2d\Omega^2,
\end{equation}
where
\begin{equation}
\begin{split}
    & X^{\pm}=\pm\frac{1}{\kappa}e^{\pm\kappa(t\pm r_*)}.
    \end{split}
\end{equation}
They relate to Kruskal coordinates via $\tau=(X^++X^-)/2$ and $\rho=(X^+-X^-)/2$.
The tortoise coordinate $-\infty<r_*<+\infty$, so $0< X^+< \infty$ and $-\infty< X^-< 0$. 
Analytically extend $X^+$ and $X^-$ to $-\infty<X^+,X^-<\infty$ subject to $r>0$, and we get an eternal KK black hole that we will work in. The Penrose diagram of a KK black hole is just as Schwarzschild's, see Fig. \ref{fig:penrose}.
The event horizon divides it into four wedges, namely the right wedge, which is the original spacetime outside the horizon, the left wedge, the future wedge and the past wedge.
And the extended coordinates $X^\pm$ relate to $(t,r_*)$ coordinates of the two-sided black hole as
\begin{equation}
\begin{split}
    & X^{\pm}=\pm\frac{1}{\kappa}e^{\pm\kappa(t\pm r_*)},\ \text{for the right wedge}\\
    & X^{\pm}=\frac{1}{\kappa}e^{\pm\kappa(t\pm r_*)},\ \text{for the left wedge}\\
    & X^{\pm}=\mp\frac{1}{\kappa}e^{\pm\kappa(t\pm r_*)},\ \text{for the future wedge}\\
    &X^{\pm}=-\frac{1}{\kappa}e^{\pm\kappa(t\pm r_*)},\ \text{for the past wedge}
    \end{split}
\end{equation}
The extension of geometry implies the existence of entanglement between the two sides, which is the spirit of "ER=EPR" \cite{Maldacena:2013xja}.
In addition, since we are considering the BH in equilibrium to a bath, the extension can be thought of as introducing a thermofield double to purify the original thermal state.
The initial state at $t=0$, known as thermofield double state (TFD), is prepared by Euclidean path integral over half the Euclidean disk illustrated by the shaded region in Fig. \ref{fig:penrose}.

\section{Entanglement entropy}
\label{sec:entropy}
In this section, we will compute the entanglement entropy for the Hawking radiation emitted by the KK black hole. 
Two cases, without island and with just one island, are considered. According to gravitational path integral, all possible configurations should be included. 
However, inspired by other works \cite{Hashimoto:2020cas,Wang:2021woy,Kim:2021gzd,Karananas:2020fwx}, we assume the single island geometry is the dominant one, and we shall also see that this is enough to reproduce the Page curve.

To avoid IR divergence, we set finite cutoff surfaces $b_{\pm}$ well beyond the black hole horizon on both left and right wedges to bipartite the two-sided geometry.
Their coordinates are $b_+=(t_b,b)$ and $b_-=(-t_b,b)$ respectively with $b\gg r_h$. 
We define the radiation region $\mathcal{R}=[b_-,\infty_-)_L\cup[b_+,\infty_+)_R$, where the subscripts $L$ and $R$ indicate they are in left and right Rindler wedges and $\infty_\pm=(\pm t_b,\infty)$.
We assume that the distance between points of our interest is much larger than the length scale of their size so that we can only consider the s-wave contribution and ignore the $d \Omega^2$ term in the metric and use 2D CFT results. 

We would like to compute the entanglement entropy of $\cal R$ with and without island
\begin{equation}
    I=[a_-,a_+],\quad a_{\pm}=(\pm t_a,a).
\end{equation}
Since the total two-sided system is in a pure state $|\mathrm{TFD}\rangle$, the entanglement entropy of $\cal R$ is equal to that of its complement $S([b_-,b_+])$. 
This is the configuration of no island.
In CFT$_2$, the entanglement entropy for one interval $[b_-,b_+]$ is given by \cite{Almheiri:2019psf}
\begin{equation}\label{entropyCFT1}
    S_{\mathrm{mat}}([b_-,b_+])=\frac{c}{3}\log d(b_-,b_+),
\end{equation}
where $d(x,y)$ is given as follows
\begin{equation}
    d^2(x,y)=|\Phi(x)\Phi(y)(U(x)-U(y))(\bar U(x)-\bar U(y))|
\end{equation}
for Euclidean metric with the form $d s^2=\Phi^2d Ud \bar U$.
Eq. \eqref{entropyCFT1} can be derived from the Weyl transformation of twist operators.
The local UV-divergent part in entanglement entropy can be absorbed into the renormalized Newton constant \cite{Susskind:1994sm} and the finite contribution is then $\eqref{entropyCFT1}$. 
The generalized entropy is then
\begin{equation}
    S_{\mathrm{gen}}=S_0+S_{\mathrm{mat}}([b_-,b_+]),
\end{equation}
where $S_0$ are terms irrelevant to $t_b$ including the area term $2\pi b^2H^{1/2}(b)/G_N$.

As for the case of an island, the entropy is given by the formula for two intervals
\begin{equation}\label{entropyCFT2}
  S_{\mathrm{mat}}(\mathcal{R}\cup I)=\frac{c}{3}\log\left[
    \frac{d(a_+,a_-)d(b_+,b_-)d(a_+,b_+)d(a_-,b_-)}{d(a_+,b_-)d(a_-,b_+)}
    \right].
\end{equation}
And we obtain the generalized entropy
\begin{equation}\label{sgen}
    S_{\mathrm{gen}}=\frac{\mathrm{Area}(\partial I)}{4G_{N}}+S_{\mathrm{mat}}(\mathcal{R}\cup I)+S_0.
\end{equation}
We should maximize $S_{\mathrm{gen}}$ in time direction $t=t_a$, but minimize it in radial direction $r=a$. 
The extremal condition fixes the location of island boundary $a_\pm$. 
And $a_\pm$ in turn give the minimum of $S_{\mathrm{gen}}$, which is expected to be finite to follow the Page curve. 
Since $S_0$ has nothing to do with the location of $\partial I$ and thus is not important in the analysis, we will drop it in the followings.

\subsection{Entanglement entropy without island}
\label{sec:entropy_noisland}
We evaluate the entanglement entropy in absence of island.
By using \eqref{entropyCFT1}, we obtain the expression for the entropy without island,
\begin{equation}\label{Snoisland}
    S_{\mathrm{mat}}([b_-,b_+])=\frac{c}{6}\log\left|
    \frac{4W(b)}{\kappa^2}\cosh^2\kappa t_b
    \right|.
\end{equation}
At early times when $t_b\ll 1/\kappa$, the entropy is approximately given by
\begin{equation}\label{S_noisland_early}
    S_{\mathrm{gen}}\simeq\frac{c}{6}\log\left|
    \frac{4W(b)}{\kappa^2}
    \right|+\frac{c}{6}\kappa^2 t_b^2.
\end{equation}
We see that the first term is the entanglement entropy for $\mathcal{R}$ in initial state up to $S_0$.
The entanglement entropy grows as $t_b^2$ from that. 

At the late-time limit $t_b\gg 1/\kappa$, this entropy is linear in time:
\begin{equation}\label{latetime_noisland}
    S_{\mathrm{gen}}\simeq \frac{c}{3}\kappa t_b,
\end{equation}
which implies that $S_{\mathrm{gen}}$ will eventually exceed the Bekenstein-Hawking entropy $2S_{\mathrm{BH}}=2\pi r_h^2H(r_h)^{1/2}/G_N$ without a bound, which indicates that the black hole spacetime evolves from a pure state to a mix state and thus the information losses.  


Notice that the above conclusions are universal for all spherically symmetric black holes even if they are in higher dimensions, because we did not use the specific form of metric. 

\subsection{Entanglement entropy with an island}

Now we will consider the case in which a single island is presented.
 Using \eqref{entropyCFT2} and \eqref{kruskal}, after some algebra, we arrive at
\begin{align}\label{Sisland}
    S_{\mathrm{gen}}=&\frac{2\pi a^2H(a)^{1/2}}{G_N}+\frac{c}{6}\log\left|
    \frac{16 W(a)W(b)}{\kappa^4}\cosh^2\kappa t_a\cosh^2\kappa t_b
    \right|\notag\\
    &+\frac{c}{3}\log\left|
    \frac{\cosh\kappa(r_*(a)-r_*(b))-\cosh\kappa(t_a-t_b)}{\cosh\kappa(r_*(a)-r_*(b))+\cosh\kappa(t_a+t_b)}
    \right|
\end{align}
for island outside the horizon, and 
\begin{align}\label{Sisland1}
    S_{\mathrm{gen}}=&\frac{2\pi a^2H(a)^{1/2}}{G_N}+\frac{c}{6}\log\left|
    \frac{16 W(a)W(b)}{\kappa^4}\sinh^2\kappa t_a\cosh^2\kappa t_b
    \right|\notag\\
    &+\frac{c}{3}\log\left|
    \frac{\sinh\kappa(r_*(a)-r_*(b))+\sinh\kappa(t_a-t_b)}{\sinh\kappa(r_*(a)-r_*(b))-\sinh\kappa(t_a+t_b)}
    \right|
\end{align}
for island inside the horizon.
Note that in deriving \eqref{Snoisland}, \eqref{Sisland} and \eqref{Sisland1}, we did not resort to the specific form of the metric, i.e. $W(r)$. 
Thus these results also apply to all spherically symmetric black holes with $g_{00}=-W(r)$. 
For black holes with $g_{00}g_{11}\neq -1$, Eqs. \eqref{Snoisland}, \eqref{Sisland} and \eqref{Sisland1} also work despite the different definitions of tortoise coordinate $r_*$. 

At the early stage where $t_b\ll r_h$, there is no extreme point for the generalized entropy by varying $t_a$ and $a$ no matter the island is inside or outside the horizon.
Thus there is no island configuration at early times.

Now we discuss the late-time behavior of island. The following assumptions are appropriately made for late times, $t_a,t_b\gg r_*(b)-r_*(a)\gg 1/\kappa$.
They further lead to
\begin{equation}
    \cosh \kappa t\simeq \frac{1}{2}e^{\kappa t}\gg \cosh \kappa(r_*(b)-r_*(a))\simeq \frac{1}{2}e^{\kappa(r_*(b)-r_*(a))}\gg 1.
\end{equation}
And we could write $S_{\mathrm{gen}}$ as
\begin{align}
S_{\mathrm{gen}}\simeq& \frac{2\pi a^2H^{1/2}(a)}{G_N}+\frac{c}{6}\log\left|
\frac{16W(a)W(b)}{\kappa^4}\left(\frac{e^{\kappa(t_a+t_b)}}{4}\right)^2
\right|+\frac{c}{3}\log\left[
\frac{\frac12e^{\kappa(r_*(b)-r_*(a))}-\cosh\kappa(t_a-t_b)}{\frac12e^{\kappa(r_*(b)-r_*(a))}+\frac{1}{2}e^{\kappa(t_a+t_b)}}
\right]\notag\\
\simeq&\frac{2\pi a^2H^{1/2}(a)}{G_N}+\frac{c}{6}\log\left|
\frac{W(a)W(b)}{\kappa^4}e^{2\kappa(r_*(b)-r_*(a))}
\right|-\frac{2c}{3}e^{\kappa(r_*(a)-r_*(b))}\cosh\kappa(t_a-t_b).\label{sgen_latetime}
\end{align}
The above expression is just the sum of the entanglement entropies of the two intervals $[a ,b]$, because at late times, the proper distance between the two intervals is very large \cite{Almheiri:2019yqk,Hashimoto:2020cas}.
It is clear that $t_a=t_b$ maximize the generalized entropy.
Actually, here we can see that the time dependence is eliminated when we set $t_a=t_b$, and the entropy will approach a constant at late times.
The entanglement entropy \eqref{sgen_latetime} will dominate over that without an island \eqref{Snoisland}, leading to a finite entanglement entropy of Hawking radiation.
This is again universal for all spherically symmetric black holes since the specific form of $W(r)$ is not yet incorporated. 

Next consider the extremal condition $\partial_aS_{\mathrm{gen}}=0$ under $t_a=t_b$.
The equation is given by
\begin{equation}\label{psgen=0}
    \frac{4a\pi \sqrt{H(a)}}{cG}+\frac{a^2\pi H'(a)}{cG\sqrt{H(a)}}-\frac13\kappa r_*'(a)+\frac{W'(a)}{6W(a)}-\frac23\kappa r_*'(a)e^{\kappa(r_*(a)-r_*(b))}=0.
\end{equation}
Numerically, it is directly to find that the minimum of $S_{\rm gen}$ is slightly outside the horizon.
Thus, we use the ansatz that $a= r_h+\epsilon$ with $\epsilon\ll r_h$.
The tortoise coordinate $r_*(a)$ is pathological at $a=r_h$, rendering the analysis difficult. 
This pathology should be fake because $r_h$ is really a coordinate singularity. 
The key point is that the factor $e^{-2\kappa r_*(a)}W(a)$ is finite at the horizon.
Actually, this must be satisfied generically.
Near the horizon, $r_*(r)$ behaves as
\begin{equation}
    r_*(r)\approx r_h\sqrt{H(r_h)}\left(
    1+\log\left|
    \frac{r-r_h}{4r_h}\frac{1}{1+r_h/Q}
    \right|
    \right)+(Q+2r_h)\mathrm{arcsinh}(\sqrt{r_h/Q})+\cdots
\end{equation}
where $"\cdots"$ denotes higher-order terms in $r-r_h$.
Then it is easy to see that this factor is finite at horizon
\begin{equation}
    e^{-2\kappa r_*(r)}W(r)|_{r\gtrsim r_h}=4 e^{-2\kappa C-1}\frac{r_h}{r}\frac{1+r_h/Q}{\sqrt{1+Q/r}},
\end{equation}
recalling that $\kappa=1/2r_h\sqrt{H(r_h)}$. 
We have defined $C\equiv(Q+2r_h){\rm arcsinh}(\sqrt{r_h/Q})$.

Now we substitute our ansatz $a=r_h+\epsilon$ with $\epsilon\ll r_h$ in \eqref{psgen=0}.
We assume that $cG_N/r_h^2\ll 1$.
This is reasonable as you can see that $r_h^2/G_N\sim (r_h/\ell_{\mathrm{P}})^2$ is just of order Bekenstein-Hawking entropy for a black hole $S_{\mathrm{BH}}$, which typically is extremely large. For example, a proton-sized BH has $S_{\mathrm{BH}}\sim 10^{40}$.
And the central charge $c$ will not be considerably large\footnote{In \cite{Almheiri:2019qdq}, the large $c$ approximation is taken to use the quantum expectation of stress tensor in classical equation of gravity. 
In the present paper, the situation is understood as we take $1\ll c\ll r_h^2/G_N$ since the number of fields is not very large \cite{Hashimoto:2020cas}.}.
This implies that the leading terms in $cG$ and leading terms in $\epsilon$ are balanced to give the extreme point.
The resulting equation is
\begin{equation}
    \frac{\pi(3Q+4r_h)}{cG\sqrt{H(r_h)}}-\frac{\kappa}{3}e^{\kappa(C-r_*(b))+1/2}\sqrt{\frac{Q}{\epsilon}}=0
\end{equation}
The solution for $\epsilon$ easily reads
\begin{equation}\label{a_island}
    \epsilon=\frac{(cG)^2}{r_h^3}\frac{e^{1-2\kappa r_*(b)}}{144\pi^2}\frac{\delta^{1-\lambda}}{4}\left(1+\sqrt{1+\delta}\right)^{2\lambda},\quad \lambda\equiv \frac{1+\delta/2}{\sqrt{1+\delta}}\geq 1.
\end{equation}
It is easy to see that $Q$ extends the island compared with the Schwarzschild case.
In the case of $Q\ll r_h$, this is more manifest. Using $Q\ll r_h$ and \eqref{rs1stQ}, we have
\begin{equation}
    a=r_h+\frac{(cG)^2}{r_h^3}\frac{e^{2\kappa(r_h-b)}}{144\pi^2(b/r_h-1)}\frac{1}{\sqrt{H(r_h)}}.
\end{equation}
In the limit $Q\rightarrow 0$, we recover the result in \cite{Hashimoto:2020cas} for the Schwarzschild black hole.
The deviation of  KK black holes from Schwarzschild black holes to the linear order in $Q$ can be written as
\begin{equation}\label{epsilonQ}
    \epsilon(Q)\simeq \epsilon(0)+\delta \kappa_0b\epsilon(0)+\mathcal{O}(Q^2),
\end{equation}
where we used $b\gg r_h$, owing to which the impact from surface gravity overwhelms that from the change of area entropy.
The subscript "0" indicates that the quantity is evaluated as $Q=0$.

Eq. \eqref{a_island} shows that at late times, the boundaries of island $a_+$ and $a_-$ locate slightly outside the horizon with an amount $\epsilon\sim (cG_N)^2e^{-2\kappa r_*(b)}/r_h^3\ll \ell_{\mathrm{P}}(\ell_{\mathrm{P}}/r_h)^3 \ll \ell_{\rm P}$.
This means that the boundary of island is within a proper distance $ds\lesssim \ell_{\rm P}(\ell_{\rm P}/r_h)\ll\ell_{\rm P}$ from the horizon, which is far smaller than the Planck length.
This is reminiscent of the \emph{trans-Planckian} problem \cite{tHooft:1984kcu}.
And because of the smallness of $\epsilon$, we can safely keep only the area term in \eqref{sgen_latetime} for the late-time entropy
\begin{equation}\label{Sgen_result}
    S_{\rm gen}=2S_{\rm BH}+\cdots.
\end{equation}

\section{Higher dimensional Kaluza-Klein black holes}
\label{sec:DdimKKBH}
In this section, we discuss the entanglement entropy for Kaluza-Klein black holes in higher dimensions.
For simplicity, we take $Q$ to be small and keep only the linear order in $Q$.
The general metric takes the form \cite{Liu:2012jra,Wu:2011zzh,Ma:2020kwc}
\begin{equation}
    d s^2=-H(r)^{-\frac{D-3}{D-2}}f(r)d t^2+H(r)^{\frac{1}{D-2}}
    \frac{d r^2}{f(r)}+H(r)^{\frac{1}{D-2}}r^2d\Omega_{D-2}^2,
\end{equation}
where
\begin{equation}
    H(r)=1+\frac{Q}{r^{D-3}},\quad f(r)=1-\frac{r_h^{D-3}}{r^{D-3}}.
\end{equation}
We can define a new coordinate such that 
\begin{equation}
    d s^2=-W(r)(d t^2-d r_*^2),
\end{equation}
where we have suppressed $d\Omega^2$ and defined
\begin{equation}
    d r_*=\frac{d r}{q(r)},\quad W(r)=H(r)^{-\frac{D-3}{D-2}}f(r),
\end{equation}
in which $q(r)$ is defined by
\begin{equation}
    q(r)\equiv \frac{f(r)}{\sqrt{H(r)}}.
\end{equation}
After integration, the tortoise coordinate $r_*(r)$ is given by the Appell series $F_1$ as
\begin{equation}\label{appell}
    r_*(r)=r~ {F}_1\left(\frac{1}{3-D};-\frac12,1;\frac{D-4}{D-3};-\frac{Q}{r^{D-3}},\frac{r_h^{D-3}}{r^{D-3}}\right).
\end{equation}
Note that this formula does not apply to 4D.
Due to the Appell series $F_1$ in \eqref{appell}, it is hard to work $e^{-2\kappa r_*(a)}W(a)$ out analytically when $a\rightarrow r_h$.
But it is tractable when $\delta\equiv Q/r_h^{D-3}\ll 1$.
To linear order in $Q$,
\begin{align}
    W(r)&=\left(
    1-\frac{D-3}{D-2}\frac{Q}{r^{D-3}}
    \right)f(r),\\
    q(r)&=f(r)\left(
    1-\frac{Q}{2r^{D-3}}
    \right)\equiv f(r)A(r),\\ r_*(r)&=r~_2F_1\left(
    1,\frac{1}{3-D},\frac{4-D}{3-D},\frac{r_h^{D-3}}{r^{D-3}}
    \right)+\frac{r\delta}{2} \left(
    _2F_1\left(
    1,\frac{1}{3-D},\frac{4-D}{3-D},\frac{r_h^{D-3}}{r^{D-3}}
    \right)-1
    \right),\label{Dr*}
 \end{align}
where $_2F_1$ is the hypergeometric function.
For convenience, we will instead write it as $_2F_1(r_h^{D-3}/r^{D-3})$, where the first three arguments are understood to be $1$, $1/(3-D)$ and $1+1/(3-D)$.
The surface gravity on the horizon is given by
\begin{equation}
\begin{split}
       \kappa=&\frac{f'(r_h)}{2\sqrt{H(r_h)}}\\
       \simeq&\frac{D-3}{2r_h}\left(1-\frac{\delta}{2}\right).
\end{split}
\end{equation}

Since the discussion of no-island case in Sec. \ref{sec:entropy_noisland} is general, we investigate only the island configuration in higher dimensions in the followings.
And because there is no early-time island, we only present the late-time result.

In higher dimensions, $S_{\mathrm{gen}}$ takes the same form as \eqref{sgen_latetime}. Similarly, we get divergence at $a=r_h$.
To cure the pathology, we write down explicitly
\begin{equation}
    e^{2\kappa r_*(a)}=\exp\left[
     \frac{(D-3)a}{r_h}~_2F_1\left(\frac{r_h^{D-3}}{a^{D-3}}\right)-\frac{D-3}{2}\frac{a}{r_h}\delta
    \right],
\end{equation}
and find that 
\begin{align}\label{gd}
    \lim_{a\rightarrow r_h}\frac{
    \exp\left[
    \frac{(D-3)a}{r_h}~_2F_1\left(\frac{r_h^{D-3}}{a^{D-3}}\right)
    \right]
    }{1-\frac{r_h^{D-3}}{a^{D-3}}}=&\exp\left[
    \gamma+\psi\left(\frac{1}{3-D}\right)
    \right]\notag\\
    =& \exp\left[\mathcal H\left(
    -\frac{D-2}{D-3}
    \right)\right]\notag\\
    \equiv& g_D,
\end{align}
where $\gamma\simeq 0.57721$ is the Euler's gamma, $\psi(z)$ is the polygamma function, and $\mathcal{H}(x)$ is the Harmonic number.
In this regard, we have
\begin{equation}
    \lim_{a\rightarrow r_h}f(a)e^{-2\kappa r_*(a)}\simeq
    \frac{\exp\left[
    \frac{D-3}{2}\frac{a}{r_h}\delta
    \right]}{g_D}
\end{equation}
Now we eliminate the pathological behavior at $r_h$, and $S_{\mathrm{gen}}$ becomes
\begin{equation}
    \begin{split}
    S_{\mathrm{gen}}=&\frac{a^{D-2}\Omega_{D-2}H^{1/(D-2)}(a)}{2G_N}+\frac{c}{6}\left[
    \log\left[\frac{W(b)}{\kappa^4g_D}
    \left(1-\frac{D-3}{D-2}\frac{Q}{a^{D-3}}
    \right)
    \right]+2\kappa r_*(b)+\frac{D-3}{2}\frac{a}{r_h}\delta
    \right]\\
    &-\frac{2c}{3}\sqrt{f(a)g_D}\exp\left(-\kappa r_*(b)-\frac{D-3}{4}\frac{a}{r_h}\delta\right)\cosh\kappa(t_a-t_b).
    \end{split}
\end{equation}
We can solve the equation $\partial_a S_{\mathrm{gen}}$ by plugging $a=r_h+\epsilon$ for $\epsilon$. The solution to the first order approximation is
\begin{equation}
    \begin{split}
            a=&r_h+\frac{(4cG_N)^2(D-2)^2(D-3)g_Dr_h^{5+2D}\exp\left[
    -\frac{\delta}{2}(D-3)-2\kappa r_*(b)
    \right]}{\left[
    cG_N\delta r_h^{2+D}(D-3)(3D-8)+6r_h^{2D}(\delta+(D-2)^2)\Omega_{D-2}
    \right]^2}\\
    \simeq&r_h+\frac{(2cG_N)^2}{r_h^{2D-5}}\frac{(D-3)g_De^{
    -2\kappa r_*(b)
    }}{\left[
    3(D-2)\Omega_{D-2}
    \right]^2}\left(
    1-\frac{\delta}{2}(D-3)
    \right).
    \end{split}
\end{equation}
Again, $cG_N/r_h^{D-2}\ll 1$ is assumed.
Though Eq. \eqref{gd} does not apply to 4 dimension, we can still recover \eqref{a_island} by setting $D=4$ and $g_4=1$.
The effect of the charge $Q$ resides in the area of the horizon and the surface gravity $\kappa\leq \kappa_0$. The existence of $Q$ will increase the late-time island by a very tiny amount. 
The extremized entropy is given by
\begin{equation}\label{S_island_latetime_D}
    \begin{split}
    S_{\mathrm{gen}}=&
    \frac{\Omega_{D-2}r_h^{D-2}H^{1/(D-2)}(r_h)}{2G_N}+
    \frac{c}{6}\left[
    \log\left(\frac{W(b)}{g_D\kappa^4}\right)+2\kappa r_*(b)
    \right]\\&+\frac23\exp\left(
    -\kappa r_*(b)
    \right)\left[
    (D-3)g_D\frac{\epsilon}{r_h}
    \right]^{1/2}+\mathcal{O}(\epsilon).
    \end{split}
\end{equation}
Again, we see that the main contribution to generalized entropy for Kaluza-Klein black holes comes from the Bekenstein-Hawking entropy that is proportional to the area of horizon. 


\section{Page curve and scrambling time}
\label{sec:page_scrambling}
In this section, we discuss the Page curve reproduced from our calculation. We also evaluate the Page time and scrambling time for KK black holes and discuss the impact from $Q$.

\begin{figure}
    \centering
    \includegraphics[scale=0.15]{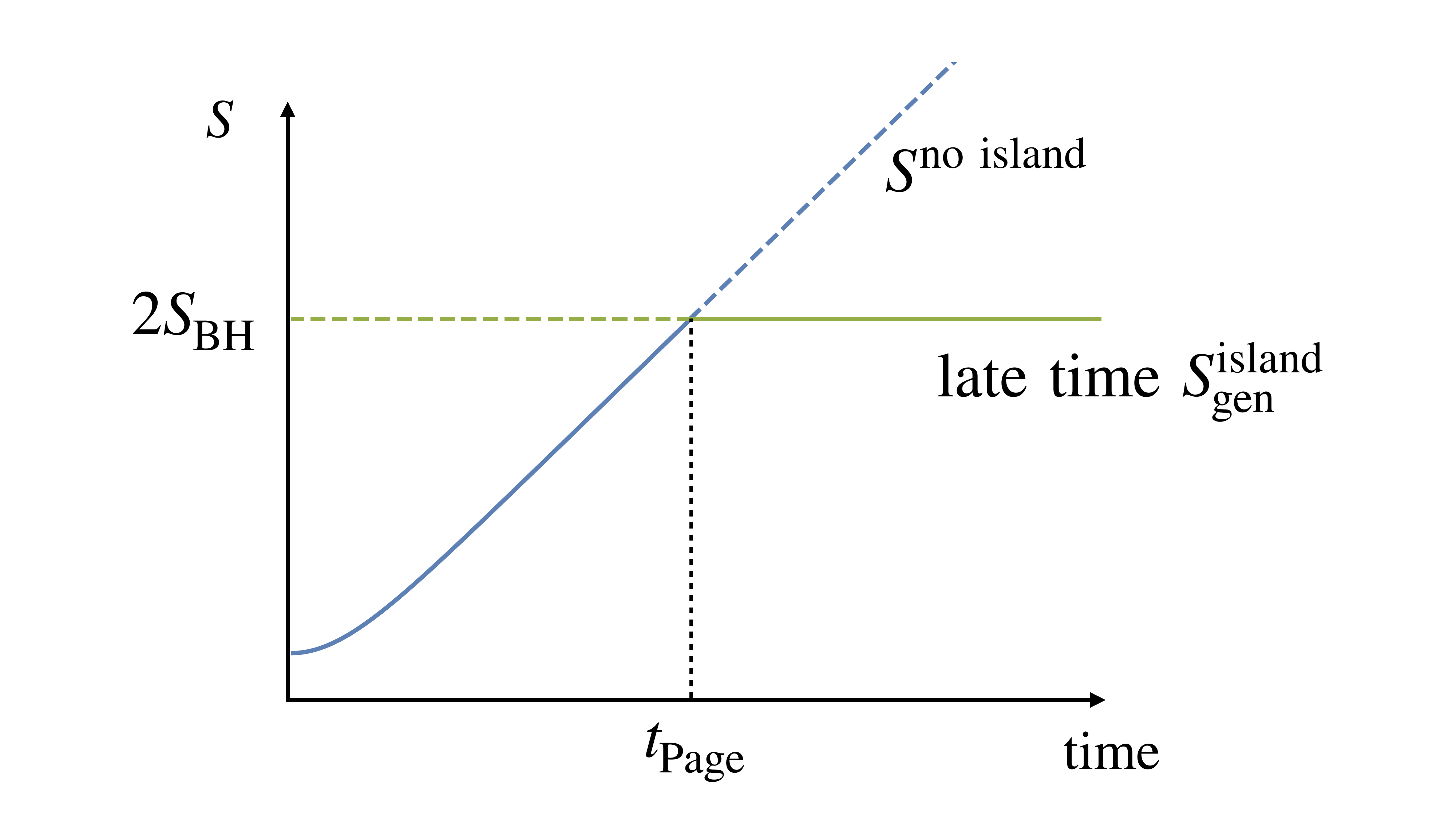}
    \caption{An illustration of the time evolution of entropy of Hawking radiation for Kaluza-Klein black holes.  The blue line represents the entropy without island while the horizontal green line stands for the late-time generalized entropy with an island. The solid part of these curves are understood as the entanglement entropy for Hawking radiation according to QES prescription. A phase transition happens at the Page time $t_{\mathrm{Page}}$.}
    \label{fig:page_curve}
\end{figure}

As our calculation suggests, at early times, there is no island, leading to the conclusion that at early times the entanglement entropy of Hawking radiation is given by the result without island \eqref{Snoisland}.
As time goes on, $S^{\mathrm{no~island}}$ grows almost linearly without a bound, and at some late time the island appears slightly outside the horizon. 
The linearly growing entropy will eventually exceed $S_{\mathrm{gen}}^{\mathrm{island}}$.
From \eqref{Sgen_result} and \eqref{S_island_latetime_D}, we see that the main contribution to $S_{\mathrm{gen}}^{\mathrm{island}}$ is from the area term, which is twice the Bekenstein-Hawking entropy for a black hole,
\begin{equation}
    S_{\mathrm{gen}}^{\mathrm{island}}(t_b\rightarrow\infty)=2S_{\mathrm{BH}}+\cdots.
\end{equation}
Therefore at late times, the true entropy is $S(\mathcal{R})\approx 2S_{\mathrm{BH}}$. So the turning point, namely the Page time, is given by
\begin{equation}
    t_{\mathrm{Page}}\approx\frac{3r_h^{D-1}\Omega_{D-2}}{cG_N(D-3)}\left(
    1+\frac{Q}{r_h^{D-3}}
    \right)^{\frac{D}{2(D-2)}}
\end{equation}
We draw an illustration of the time evolution of entanglement entropy of Hawking radiation, see Fig. \ref{fig:page_curve}, which is the Page curve of an eternal black hole.
Compared to the Page time of a Schwarzschild black hole that amounts to setting $Q=0$, the Page time of a Kaluza-Klein black hole is prolonged by a factor $H(r_h)^{D/(2D-4)}$.
In terms of black hole temperature $T=\kappa/2\pi$, the Page time can be written as
\begin{equation}\label{pagetime}
    t_{\mathrm{Page}}=\frac{3}{\pi c}\frac{S_{\mathrm{BH}}}{T}.
\end{equation}
The delay of the Page time in KK black holes can be understood in the following way. 
The charge $Q$ reduces the surface gravity as well as the Hawking temperature. 
The lower the temperature is, the weaker the emission and absorption will be, and more slowly the entanglement entropy will grow. 
But $Q$ has almost no effect on the late-time location of $a$, so $S_{\mathrm{gen}}^{\mathrm{island}}$ is invariant.
Thus it takes a longer time for no-island entanglement entropy to reach $2S_{\mathrm{BH}}$.
If we consider the backreaction of Hawking radiation instead of an eternal black hole, the black hole will eventually evaporate. This process leads to a decreasing in $S_{\mathrm{gen}}^{\mathrm{island}}$ and thus in $S(\mathcal{R})$.
Besides, the life time of an evaporating black hole is at the same order of Page time \cite{Hashimoto:2020cas}.
That is to say, $t_{\mathrm{Page}}\sim t_{\mathrm{eva}}\sim S_{\mathrm{BH},i}/T_{,i}$, where the subscript $i$ means initial value.

Now we discuss the scrambling time $t_{\mathrm{scr}}$ \cite{Hayden:2007cs} in 4-dimensional case for simplicity. Suppose we throw a photon at $b_+$ towards the black hole. 
In a finite time, it will reach the island since it's located outside the horizon. The island is in the entanglement wedge of radiation, and the information about this photon could be decoded by the outside Hawking radiation.
So the scrambling time is
\begin{align}
    t_{\rm scr}=&r_*(b)-r_*(a)
    \notag\\
    =&b\sqrt{H(b)}-a\sqrt{H(a)}+(Q+2r_h)\log\left[
\frac{\sqrt{b}(1+\sqrt{H(b)})}{\sqrt{a}(1+\sqrt{H(a)})}\right]\notag\\
&+r_h\sqrt{H(r_h)}\log\left[
\frac{b-r_h}{a-r_h}\left(\frac{\sqrt{(a+Q)r_h}+\sqrt{(Q+r_h)a}}{\sqrt{(b+Q)r_h}+\sqrt{(Q+r_h)b}}\right)^2
\right]\notag\\
\simeq & r_h\sqrt{H(r_h)}\log\left[
\frac{b-r_h}{a-r_h}
\right]\notag\\
\simeq&r_h\sqrt{H(r_h)}\log S_{\rm BH},\label{tscr}
\end{align}
where we have use $r_h^2/G_N\sim S_{\rm BH}$.
Eq. \eqref{tscr} is in consistent with \cite{Hashimoto:2020cas,Wang:2021woy,Hayden:2007cs,Harlow2016,Sekino:2008he}. It seems $Q$ prolongs the scrambling time by a factor $\sqrt{H(r_h)}$.
We can rewrite $t_{\rm scr}$ using temperature $T$ as
\begin{equation}
    t_{\rm scr}\sim \frac{1}{T}\log S_{\rm BH}\sim r_h\log \frac{r_h}{\ell_{\rm P}},
\end{equation}
and in terms of $t_{\rm Page}$
\begin{equation} 
    t_{\mathrm{scr}}\sim \frac{1}{T}\log Tt_{\mathrm{Page}}\ll t_{\mathrm{Page}}.
\end{equation}
We see that the scrambling time $t_{\rm scr}$ is small compared with the Page time.

\section{Conclusion and discussion}
\label{sec:ccs}

In this paper, we studied the entanglement island in the scenario of non-rotating Kaluza-Klein black holes. 
KK black holes include the Schwarzschild black holes in the limit of vanishing charge $Q=0$. We can recover the results in Schwarzschild black holes \cite{Hashimoto:2020cas} by sending $Q\rightarrow 0$.

For no-island configuration, the generalized entropy for Hawking radiation grows first in $t_b^2$ and then linearly without a bound \eqref{Snoisland}.
This is the information problem for an eternal black hole.
We then consider the island configuration.
For island outside the horizon, the generalized entropy is given by \eqref{Sisland}, and for island inside the horizon, the generalized entropy is given by \eqref{Sisland1}.
At early times, there is no extremal point for $S_{\rm gen}$ by varying $a$ and $t_a$.
Thus there is no island at early times.
At late times, island appears with the boundary slightly outside the horizon $a\gtrsim r_h$.
Then island saves the day for information paradox in KK black hole in sense that the Page curve is reproduced.

We investigated the impact of the charge $Q$ on the behavior of the island at late times.
According to our calculation, the location of $\partial I$ is outside the horizon by $\epsilon\sim (cG_N)^2/r_h^3\lesssim \ell_{\mathrm{P}}(\ell_{\mathrm{P}}/r_h)^3\ll r_h$.
It gives a constant entanglement entropy that is twice the Bekenstein-Hawking entropy $S_{\mathrm{gen}}^{\mathrm{island}}\simeq 2S_{\mathrm{BH}}$.
The linearly growing $S^{\mathrm{no~island}}$ will eventually exceed $2S_{\mathrm{BH}}$ at the Page time given in \eqref{pagetime}. 
So the Page curve for an eternal black hole is reproduced as shown in Fig. \ref{fig:page_curve}. 
In fact, for all spherically symmetric black holes, a similar behavior for entanglement entropy is expected if the island rule is applied, as in \cite{Almheiri:2019yqk,Hashimoto:2020cas,Wang:2021woy,Kim:2021gzd}. 
This is because we already see this behavior in \eqref{latetime_noisland} and \eqref{sgen_latetime} before the involvement of explicit form of spacetime metric.

In addition, compared with Schwarzschild black holes, $Q$ will enlarge the island, see \eqref{a_island}, but its boundary is still very close to the horizon $\epsilon\ll r_h$.
To the linear order in $Q$, $\Delta a=\delta \kappa_0b\epsilon(0)+\mathcal O(Q^2)$, see \eqref{epsilonQ}.
The correction to Bekenstein-Hawking entropy is too small to be considered in the discussion of Page time. 
We then find that $Q$ will delay the Page time by a factor $(1+Q/r_h^{D-3})^{D/2(D-2)}$.
In 4-dimension, the scrambling time is prolonged by a factor $(1+Q/r_h)^{1/2}$. 
We also generalize the computation of entanglement entropy using island formula to higher dimensions $D\geq 5$ with $Q\ll r_h$.
At late times, it turns out $a\sim r_h+\ell_{\mathrm{P}}(\ell_{\mathrm{P}}/r_h)^{2D-5}$, where $\ell_{\mathrm{P}}=G_{N}^{1/(D-2)}$. 
The late-time generalized entropy will also be twice the Bekenstein-Hawking area entropy $2S_{\mathrm{BH}}$.

In a word, $Q$ does enlarge the island, delay the Page time and prolong the scrambling time, but the island rule still leads to a Page curve for KK black holes, implying a unitary evolution.

However, a problem is that the boundary of island is close to the horizon less than the Planck length scale.
The quantum effect of gravity seems to play an important role at this scale.
The precise location of the entanglement wedge, after taking this into account, remains to be explored.

\begin{acknowledgements}
We thank Jiang Long, Wuzhong Guo and Liang Ma for useful discussion. 
This research is supported in part by the National Natural Science Foundation of China under Grant No. 11875136,
and the Major Program of the National Natural Science Foundation of China under Grant No. 11690021.
\end{acknowledgements}

\appendix

\providecommand{\href}[2]{#2}\begingroup\raggedright\endgroup

\end{document}